\documentclass[prd,twocolumn,showpacs,superscriptaddress,nofootinbib,floatfix,showkeys,10pt]{revtex4-1}
\usepackage{graphicx}
\usepackage{amsmath}
\usepackage{bm}
\usepackage{yhmath}
\usepackage{mathtools}
\usepackage{wasysym}
\usepackage[colorlinks,citecolor=blue,urlcolor=blue,linkcolor=blue]{hyperref}
\usepackage{subfigure}
\usepackage{color}
\usepackage{cases}
\usepackage{subfigure}
\usepackage{times}
\usepackage{dcolumn,booktabs,bm}
\usepackage{slashed}
\usepackage{amsfonts,amssymb,stmaryrd,latexsym,amsmath}
\usepackage{textcomp}
\usepackage{multirow}
\usepackage{cancel}
\usepackage{array}

\def\SU3{{\text{SU(3)}_{\rm F}}}

\begin{document}

\title{Revisit the isospin violating decays of $X(3872)$}

\author{Lu Meng}
\affiliation{Ruhr-Universit\"at Bochum, Fakult\"at f\"ur Physik und
Astronomie, Institut f\"ur Theoretische Physik II, D-44780 Bochum,
Germany }

\author{Guang-Juan Wang}
\affiliation{Advanced Science Research Center, Japan Atomic Energy
Agency, Tokai, Ibaraki, 319-1195, Japan}

\author{Bo Wang}\email{wangbo@hbu.edu.cn, corresponding author}
\affiliation{School of Physical Science and Technology, Hebei
    University, Baoding 071002, China}
\affiliation{Key Laboratory of High-precision Computation and
Application of Quantum Field Theory of Hebei Province, Baoding
071002, China}

\author{Shi-Lin Zhu}\email{zhusl@pku.edu.cn, corresponding author}
\affiliation{School of Physics and Center of High Energy Physics,
Peking University, Beijing 100871, China}

\begin{abstract}
In this work, we revisit the isospin violating decays of $X(3872)$
in a coupled-channel effective field theory. In the molecular
scheme, the $X(3872)$ is interpreted as the bound state of
$\bar{D}^{*0}D^0/\bar{D}^0D^{*0}$ and $D^{*-}D^+/D^-D^{*+}$
channels. In a cutoff-independent formalism, we relate the coupling
constants of  $X(3872)$ with the two channels to the molecular  wave
function. The isospin violating decays of $X(3872)$ are  obtained by
two equivalent approaches, which amend some deficiencies about this
issue in literature. In the quantum field theory approach, the
isospin violating decays arise from the coupling constants of
$X(3872)$ to two di-meson channels. In the quantum mechanics
approach, the isospin violating is attributed to wave functions at
the origin. We illustrate that how to cure the insufficient results
in literature. Within the comprehensive analysis, we bridge the
isospin violating decays of $X(3872)$ to its inner structure. Our
results show that the proportion of the neutral channel in $X(3872)$
is over $80\%$. As a by-product, we calculate the strong decay width
of $X(3872)\to \bar{D}^0 D^0\pi^0$ and radiative one $X(3872)\to
\bar{D}^0 D^0\gamma$. The strong decay width and radiative decay width are
about 30 keV and 10 keV, respectively, for the binding energy from
$-300$ keV to $-50$ keV.
\end{abstract}

\maketitle

\section{Introduction}~\label{sec:intro}

In 2003, the observation of $X(3872)$~\cite{Belle:2003nnu} launched
a new era of hadron spectroscopy. Amounts of candidates of exotic
hadrons (beyond the $q\bar{q}$ mesons and $qqq$ baryons) were
observed in experiments (see Refs.
\cite{Chen:2016qju,Esposito:2016noz,Lebed:2016hpi,Hosaka:2016pey,Guo:2017jvc,Olsen:2017bmm,Liu:2019zoy,Brambilla:2019esw}
for reviews). Among these exotic hadron candidates, $X(3872)$ is
undoubtedly the superstar. One of its most salient features is that
the mass coincides exactly with the $\bar{D}^{*0} D^0/\bar{D}^0
D^{*0}$ threshold as $m_{D^0}+m_{D^{*0}}-m_{X(3872)}=(0.00\pm 0.18)$
MeV~\cite{ParticleDataGroup:2020ssz}, which naturally inspired the
molecular
interpretations~\cite{Voloshin:2003nt,Swanson:2003tb,Tornqvist:2004qy,Fleming:2007rp,Liu:2008fh}.
Another important feature of $X(3872)$ is the large isospin
violating decay
patterns~\cite{Belle:2005lfc,BaBar:2010wfc,BESIII:2019qvy},
\begin{eqnarray}
&&\frac{{\cal B}[X\to J/\psi\pi^{+}\pi^{-}\pi^{0}]}{{\cal B}[X\to J/\psi\pi^{+}\pi^{-}]}=1.0\pm0.4\pm0.3\quad\text{Belle},\nonumber\\
&&\frac{{\cal B}[X\to J/\psi\omega]}{{\cal B}[X\to
J/\psi\pi^{+}\pi^{-}]}=\begin{cases}
    1.6_{-0.3}^{+0.4}\pm0.2\quad\text{BESIII,}\\
    0.7\pm0.3\quad B^{+}\text{events, BABAR,}\\
    1.7\pm1.3\quad B^{0}\text{ events, BABAR.}\nonumber
\end{cases}
\end{eqnarray}
The two features can be related to each other. The charged threshold
$D^{*-}D^+/D^-D^{*+}$ is about $8$ MeV above the neutral threshold.
Since the mass of $X(3872)$ exactly coincides with the neutral
threshold, the mass difference between two thresholds plays a
critical role on the different contents of  the di-meson components
in the molecule  which  lead to the isospin violating decays.

In the past decades, more and more refined theoretical calculations
(e.g.
Refs.~\cite{Baru:2011rs,Baru:2013rta,Zhao:2014gqa,Baru:2015tfa,Schmidt:2018vvl,Guo:2019qcn,Braaten:2020nmc,Meng:2020cbk}
) and experimental analysis (e.g.~\cite{LHCb:2015jfc,LHCb:2020xds})
were performed to investigate the nature of $X(3872)$. Its isospin
violating decays which are  sensitive to the inner structures  are
of great interest and have been investigated in different
scenarios~\cite{Tornqvist:2004qy,Suzuki:2005ha,Braaten:2005ai,Ortega:2009hj,Gamermann:2009fv,Gamermann:2009uq,Hanhart:2011tn,Li:2012cs,Takeuchi:2014rsa,Zhou:2017txt,Wu:2021udi}.
It is well accepted that the large isospin violation is induced by
the mass splitting between charged and neutral channels. However,
there are some disagreements on the specific mechanisms even in the
same picture. For example,  Refs.~\cite{Gamermann:2009fv} and
\cite{Li:2012cs} provided two different scenarios about the large
isospin violating decays of $X(3872)$ in the molecular model. In
Ref.~\cite{Gamermann:2009fv}, the large isospin violating decays
were driven by the mass difference in propagators and amplified by
the phase space. The coupling constants  of the $X(3872)$ with  the
neutral and charged di-meson channels were presumed to satisfy the
isospin symmetry. In Ref.~\cite{Li:2012cs}, the authors stressed
that the isospin violating effect arises from the wave function of
$X(3872)$ which is related to the coupling constants
~\cite{Braaten:2005ai,Gamermann:2009uq,Gamermann:2009uq,Aceti:2012dd,Sekihara:2016xnq}.
This indicated that the isospin   violating decays of $X(3872)$ were
induced by the coupling constants.  In this work, we focus on the
molecular scheme of $X(3872)$ and aim to make a comprehensive
investigation on the isospin violating decays of $X(3872)$. To this
end, we will adopt the coupled-channel effective field
theory~\cite{Cohen:2004kf,Braaten:2005ai} to clarify some
conceptional ambiguities about coupling constants, wave function and
its value at the origin. In this approach, the cutoff-dependence is
eliminated exactly. The same formalism was recently used to exploit
the structure of $T_{cc}^+$ state~\cite{Meng:2021jnw}.

The work is organized as follows. In Sec.~\ref{sec:cp}, the coupling
constants of $X(3872)$ to two di-meson channels are obtained by
solving the Lippmann-Schwinger equations (LSEs). In
Sec.~\ref{sec:wvfunc}, the Schr\"odinger equation is adopted to
discuss the relation of wave function and coupling constants. In
Sec.~\ref{sec:iso_v_decay}, the isospin violating decays of
$X(3872)$ are analyzed and some disadvantages in literature are
clarified. In Sec.~\ref{sec:decay}, the decay widths of $X(3872)\to
\bar{D}^0 D^0\pi^0$ and $X(3872)\to \bar{D}^0 D^0\gamma$  are
calculated. In Sec.~\ref{sec:sum}, the main conclusions are
summarized briefly. At last, more details about calculating
sequential decays are presented in Appendix~\ref{app:sq_decay}.

\section{Coupling constants}~\label{sec:cp}

\begin{figure}[!htp]
    \centering  \includegraphics[width=0.35\textwidth]{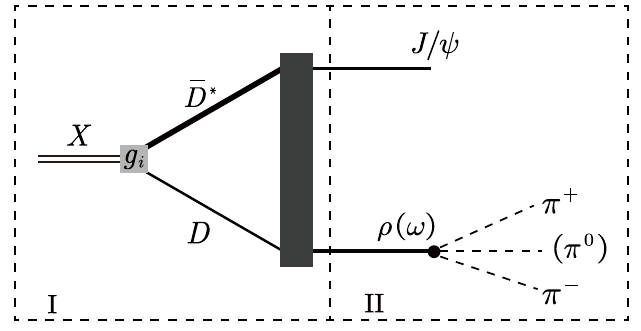}
    \caption{Isospin violating decays of $X(3872)$. }\label{fig:feynjpsipipi}
\end{figure}

 The dominant  decays of the $X(3872)$ are $D^0\bar{D}^0\pi^0$ and $D^{*0}\bar{D}^0$ with the  fractions being $(49^{+18}_{-20})\%$ and $(37\pm 9)\%$, respectively~\cite{ParticleDataGroup:2020ssz}. It couples  much stronger with the open-charmed channels (e.g. $D^{*0}\bar{D}^0$ )  than the hidden-charmed ones (e.g. $J/\psi \pi^+\pi^-$). Therefore, as an approximation, we only consider the interplay of the open-charmed channels to investigate the structure of $X(3872)$. The hidden-charmed decays can be driven by the open-charmed components of $X(3872)$ as illustrated in Fig.~\ref{fig:feynjpsipipi}.

The two closest di-meson thresholds to $X(3872)$ are the
$\bar{D}^{*0}D^0/\bar{D}^0D^{*0}$ (neutral) and
$D^{*-}D^+/D^-D^{*+}$ (charged) channels with
$m_{D^0}+m_{D^{*0}}-m_{X(3872)}=(0.00\pm 0.18)$ MeV, while the
charged one is located $8$ MeV higher.  Considering the mass
uncertainty, the $X(3872)$ might be above or below the neutral
threshold. In this work, we only focus on the bound state picture as
those in Refs.~\cite{Gamermann:2009fv,Gamermann:2009uq,Li:2012cs}.
As a bound state, one can relate the wave function of $X(3872)$ to
its coupling constants with the two di-meson channels. The similar
formalism in the work can be extended to resonance
states~\cite{Meng:2020ihj,Meng:2021rdg}.

We use a coupled-channel effective field
theory~\cite{Cohen:2004kf,Braaten:2005ai} to investigate the
$T$-matrix of neutral and charged di-meson channels. Similar method
has been used to study $X(3872)$ and $T_{cc}^+$ states in
Refs.~\cite{Braaten:2005ai,Meng:2021jnw}. Here, we introduce the
formalism briefly. The leading order contact interactions with a
hard regulator can be constructed as follows,
\begin{equation}
    V(\bm{p},\bm{p}')=\left[\begin{array}{cc}
        v_{11} & v_{12}\\
        v_{12} & v_{22}
    \end{array}\right]\Theta(\Lambda-p)\Theta(\Lambda-p'), ~\label{eq:vij}
\end{equation}
where $\Theta$ is the step function. $v_{ij}$ are the
energy-independent potentials for the di-meson channels. We use the
subscripts $1$ and $2$ to label the neutral and charged channels,
respectively. The coupled-channel LSEs can be reduced to algebraic
equations,
\begin{equation}
    T(\bm{p},\bm{p'})=t\Theta(\Lambda-p)\Theta(\Lambda-p'),\quad t=v+vGt,
\end{equation}
where $G=\text{diag}\{G_1,G_2\}$ is the diagonal matrix. The
propagator $G_i$ reads
\begin{eqnarray}
    G_{i}(E)&=&\int^{\Lambda}\frac{d^{3}\bm{p''}}{(2\pi)^{3}}\frac{1}{E-\delta_{i}-\frac{p''^{2}}{2\mu}+i\epsilon}, \nonumber\\
    &=&\frac{\mu}{\pi^{2}}\left[-\Lambda+k_{i}\arctan\frac{\Lambda}{k_{i}}\right] \nonumber\\
    &\approx&\frac{\mu}{2\pi}\left[-\frac{2}{\pi}\Lambda+k_{i}\right],\label{eq:Gi}
\end{eqnarray}
where $\delta_1=0$ and $\delta_2=
m_{D^{*+}}+m_{D^-}-m_{D^{*0}}-m_{D^0}$. $\mu$ is the reduced mass,
for which we neglect the tiny difference between the  two di-meson
channels. $k_{i}\equiv\sqrt{-2\mu(E-\delta_i)}$ is the binding
momentum. In Eq.~\eqref{eq:Gi}, the approximation is a consequence
of $|k_i|\ll \Lambda$. The $T$-matrix corresponding to physical
observables are cutoff-independent. The cutoff-dependence in the
$G_i$ is canceled out by the $v_{ij}$. To make it clear, we
introduce a new set of parameters,
\begin{eqnarray}
    \begin{cases}
        \frac{1}{b_{11}} & =\frac{2\pi}{\mu}(\frac{v_{22}}{v_{11}v_{22}-v_{12}^{2}}-G_{1})+k_{1}\\
        \frac{1}{b_{22}} & =\frac{2\pi}{\mu}(\frac{v_{11}}{v_{11}v_{22}-v_{12}^{2}}-G_{2})+k_{2}\\
        \frac{1}{b_{12}} & =\frac{2\pi}{\mu}\frac{v_{12}}{v_{11}v_{22}-v_{12}^{2}}
    \end{cases}.~\label{eq:defineb}
\end{eqnarray}
Here, the $b_{ij}$ are the cutoff independent parameters. The
solution of LSEs can be expressed as
\begin{equation}
    t=\frac{1}{D}\left[\begin{array}{cc}
        b_{11}b_{12}^{2}\left(1-b_{22}k_{2}\right) & b_{11}b_{12}b_{22}\\
        b_{11}b_{12}b_{22} & b_{12}^{2}b_{22}\left(1-b_{11}k_{1}\right)
    \end{array}\right],\label{eq:tmrx}
\end{equation}
with
$D=\frac{\mu}{2\pi}\left[b_{12}^{2}\left(b_{11}k_{1}-1\right)\left(b_{22}k_{2}-1\right)-b_{11}b_{22}\right]$.

The bound state corresponds to a pole of the $T$-matrix in the real
axis. At the pole position, the coupling constants of $X(3872)$ to
two di-meson channels can be approximated by the residues of the
$T$-matrix,
\begin{equation}
    \lim_{E\to E_{0}}(E-E_{0})t_{ij}=\lim_{E\to E_{0}}\left[\frac{d(t_{ij})^{-1}}{dE}\right]^{-1}=\frac{1}{8M_{X}^{2}\mu}g_{i}g_{j}, \label{eq:resid}
\end{equation}
where $E_0$ and $M_X$ are the binding energy and mass of $X(3872)$,
respectively. The coupling constants can be expressed in a very
simple form,
\begin{equation}
    g_{1}=\frac{4M_{X}\sqrt{\pi\kappa_{1}}}{\sqrt{\mu}}\cos\theta,\quad g_{2}=\frac{4M_{X}\sqrt{\pi\kappa_{2}}}{\sqrt{\mu}}\sin\theta,~\label{eq:g1g2}
\end{equation}
with the following parameters $\kappa_i$ and $\theta$
\begin{equation}
    \lim_{E\to E_{0}}k_{i}\equiv\kappa_{i},\quad\tan^{2}\theta\equiv\frac{b_{22}\kappa_{1}\left(b_{11}\kappa_{1}-1\right)}{b_{11}\kappa_{2}\left(b_{22}\kappa_{2}-1\right)}.\label{eq:theta}
\end{equation}
In the calculation, the $b_{22}$ is eliminated by $\lim_{E\to
E_0}D=0$. The $\theta$ is actually the mixing angle of two di-meson
components in the molecule and will be proved in the latter section.

With a special interaction $v_{11}=v_{12}=v_{22}$,  our analytical
results are in accordance with those in
Refs.~\cite{Gamermann:2009fv,Gamermann:2009uq}. What is more
important, the cutoff-dependences in the potential $v_{ij}$ and $G$ are canceled out in Eq.~\eqref{eq:defineb}. Our results are cutoff-independent, which satisfy the
renormalization group invariance ${dT\over d \Lambda}=0$. With the extra constraint,
$v_{11}=v_{12}=v_{22}$~\cite{Gamermann:2009fv,Gamermann:2009uq}, the
cutoff dependence of $T$-matrix can not be eliminated. The
assumption $v_{11}=v_{12}=v_{22}$ is motivated by vanishing
interaction in spin triplet channel, $V_{I=1}=0$. However, in the
non-perturbative renormalization formalism of LSEs, it is imprudent
to introduce the cutoff-independent interaction $V_{I=1}$.

\section{Wave functions}~\label{sec:wvfunc}
The bound state  can be investigated with Schr\"odinger equation as
well. In the coupled-channel effective field theory, the
Schr\"odinger equation reads,
\begin{equation}
    (\hat{H}_{0}+\hat{V})|\psi\rangle=E_{0}|\psi\rangle,\quad\hat{V}=\sum_{i,j}\frac{1}{(2\pi)^{3}}v_{ij}|i\rangle\langle j|,
\end{equation}
where $1/(2\pi)^{3}$ is  the  normalization factor. In the
single-channel framework, the wave function for channel $|i\rangle$
reads,
\begin{equation}
    \phi_{i}(p)=\xi_{i}\frac{\Theta(\Lambda-p)}{E_{0}-\frac{p^{2}}{2\mu}-\delta_{i}},\quad\xi_{i}^{2}\approx\frac{\kappa_{i}}{4\pi^{2}\mu^{2}} \label{eq:wvfunc},
\end{equation}
where $\xi_i$ are the normalization constant. The solution of the
coupled-channel Schr\"odinger equation can be introduced as the
combination of the single-channel wave functions,
\begin{eqnarray}
    \langle\bm{p}|\psi\rangle=c_{1}\phi_{1}(p)|1\rangle+c_{2}\phi_{2}(p)|2\rangle,
\end{eqnarray}
where $c_{1,2}^2$ are the probability of the two components  and
satisfy  $c_1^2+c_2^2=1$.  The wave function at the origin can be
obtained as
\begin{eqnarray}
    \varphi(0)  &=&\int \frac{d^{3}\bm{p}}{(2\pi)^{3/2}}\psi(\bm{p})%\nonumber\\
    = c_1 \varphi_1(0)|1\rangle+c_2 \varphi_2(0) |2\rangle \nonumber \\
    &=&(2\pi)^{3/2}\left[c_{1}\xi_{1}G_{1}|1\rangle+c_{2}\xi_{2}G_{2}|2\rangle\right]. \label{eq:wv_orgin}
\end{eqnarray}

In order to relate the coupling constants with the wave function, we
take the approximate $T$-matrix near the bound state
pole~\cite{Sekihara:2016xnq}
\begin{equation}
    T_{ij}(\bm{p},\bm{p'})\approx(2\pi)^{3}\frac{\langle\bm{p},i|\hat{V}|\psi\rangle\langle\psi|\hat{V}|\bm{p'},j\rangle}{E-E_{0}},
\end{equation}
with $\langle\bm{p},i|\hat{V}|\psi\rangle$ being calculated by
\begin{eqnarray}
    \langle\bm{p},i|\hat{V}|\psi\rangle&=&\langle p,i|\hat{H}-\hat{H}_{0}|\psi\rangle=\left[E_{0}-\frac{p^{2}}{2\mu}-\delta_{i}\right]\langle\bm{p},i|\psi\rangle \nonumber\\
    &=&c_{i}\xi_{i}\Theta(\Lambda-p).
\end{eqnarray}
The $T$-matrix reads
\begin{equation}
    t_{ij}\approx (2\pi)^{3}\frac{c_{i}c_{j}\xi_{i}\xi_{j}}{E-E_{0}}~\label{eq:tmx_wv}.
\end{equation}
With  Eqs.~\eqref{eq:resid} and ~\eqref{eq:g1g2}, one can obtain
\begin{equation}
    c_1=\cos \theta,\quad c_2=\sin\theta.
\end{equation}
Thus, we can see that the $\theta$ defined in Eq.~\eqref{eq:theta}
is in fact the mixing angle of the charged and neutral channels.

As shown by Eq.~\eqref{eq:g1g2}  and Eq.~\eqref{eq:wv_orgin}, the
coupling constants and wave function at the origin depend on the
binding energy (in $\kappa_i$) and mixing angle. The mixing angle
reflects the isospin violation effect of $X(3872)$. $\theta=\pi/4$,
and $-\pi/4$ correspond to the isospin singlet and triplet states,
respectively. $\theta=0$ stands for $X(3872)$ as a pure
$\bar{D}^{*0} D^0/\bar{D}^0 D^{*0}$ bound state.

\section{Mixing angle and isospin violating decays}~\label{sec:iso_v_decay}

The decays $X\to J/\psi\pi^{+}\pi^{-}\pi^{0}$ or $X\to
J/\psi\pi^{+}\pi^{-}$ can be decomposed into two parts, as shown in
Fig.~\ref{fig:feynjpsipipi}. We can define the  ratio as,
\begin{equation}
    R\equiv \frac{{\cal B}^{I=0}(X\to J/\psi\pi^{+}\pi^{-}\pi^{0})}{{\cal B}^{I=1}(X\to J/\psi\pi^{+}\pi^{-})}=R_{1}\times R_{2},
\end{equation}
where $R_1$ and $R_2$ represent the ratios for part  I and  part II
as illustrated in Fig.~\ref{fig:feynjpsipipi}, respectively. In
part II, the  $J/\psi \rho$ and $J/\psi \omega$ channels are both
located close to the $X(3872)$. With the widths of $\rho$ and
$\omega$ taken into account, $R_2$ will be suppressed.  Based on the
formalism of the sequential decays in Appendix~\ref{app:sq_decay},
we can estimate the $R_2$ as follows~\cite{Gamermann:2009uq},
\begin{eqnarray}
    R_{2}&=\frac{\int_{(3m_{\pi})^{2}}^{(m_{X}-m_{J/\psi})^{2}}\mathcal{S}(p_{V}^{2},m_{\omega},\Gamma_{\omega})q(m_{X},p_{V},m_{J/\psi})dp_{V}^{2}}{\int_{(2m_{\pi})^{2}}^{(m_{X}-m_{J/\psi})^{2}}\mathcal{S}(p_{V}^{2},m_{\rho},\Gamma_{\rho})q(m_{X},p_{V},m_{J/\psi})dp_{V}^{2}}\nonumber\\
    &\quad\times\frac{{\cal B}(\omega\to\pi^{+}\pi^{-}\pi^{0})}{{\cal B}(\rho^{0}\to\pi^{+}\pi^{-})},\label{eq:r2}
\end{eqnarray}
where $m_i$ and $\Gamma_i$ are the masses and widths of  the $\rho$
and $\omega$ mesons. $p_V$ is the momentum of the $\rho$ or $\omega$
in Fig.~\ref{fig:feynjpsipipi}. The $\mathcal{S}$ and $q$ are
defined as,
\begin{equation}
    \mathcal{S}(p_{V}^{2},m_{V},\Gamma_{V})=\frac{m_{V}\Gamma_{V}}{(p_{V}^{2}-m_{V}^{2})^{2}+m_{V}^{2}\Gamma_{V}^{2}},
\end{equation}
\begin{eqnarray}
    q(&&M,m_{1},m_{2})\nonumber \\
    &&=\frac{\sqrt{[M^{2}-(m_{1}-m_{2})^{2}][M^{2}-(m_{1}+m_{2})^{2}]}}{2M}.
\end{eqnarray}

 Eq.~\eqref{eq:r2} is the approximate result which neglects the polarizations of the particles and factors out  the  $\omega \to \pi^+\pi^-\pi^0 (\rho^0\to \pi^+\pi^-)$ decay widths (see Appendix~\ref{app:sq_decay} for details).  With the experimental results ${\cal B}(\rho^{0}\to\pi^{+}\pi^{-})\approx100\%$ and ${\cal B}(\omega\to\pi^{+}\pi^{-}\pi^{0})\approx 89.3\%$~\cite{ParticleDataGroup:2020ssz}, the ratio reads
\begin{equation}
    R_2\approx 0.147.\label{eq:R2}
\end{equation}
 In Ref.~\cite{Braaten:2005ai}, the authors did not take approximations mentioned above and obtained $R_2=0.087$ within a refined calculation. We will substitute both the two values of $R_2$ in the following calculation.

For part I in Fig.~\ref{fig:feynjpsipipi}, there are different
scenarios in literature~\cite{Gamermann:2009fv,Li:2012cs}. In our
framework, we connect the coupling constants with the wave function.
Therefore, we can give a more comprehensive analysis to clarify the
ambiguity in literature. We will interpret the scenarios of
Refs.~\cite{Gamermann:2009fv,Li:2012cs} with our notations and then
compare them with our approach.

We have two equivalent approaches to investigate the isospin
violating decays $X^{I=a}\to J/\psi V$ ($V$ represents the $\rho$ or
$\omega$ mesons) in part I. The di-meson state $X^{I=a}$ with
isospin $I=a,I_z=0$ is composed of the charged and neutral channels
as follows,
\begin{equation}
    |I=a\rangle=c_{a1}|1\rangle+c_{a2}|2\rangle,
\end{equation}
where $c_{ai}$ are the Clebsch-Gordan coefficients. In the quantum
field theory, the amplitude  can be calculated as
\begin{eqnarray}
    &&  {\cal A}^{I=a}[X\to J/\psi V]\nonumber\\
    \sim&&\int\frac{d^{3}q}{(2\pi)^{3}}\frac{\Theta(\Lambda-q)}{E-\frac{q^{2}}{2\mu}}c_{a1}g_{1}+\int\frac{d^{3}q}{(2\pi)^{3}}\frac{\Theta(\Lambda-q)}{E-\frac{q^{2}}{2\mu}-{\delta}}c_{a2}g_{2}\nonumber\\
    \sim&&c_{a1}{G}_{1}{g}_{1}+c_{a2}{G}_{2}{g}_{2},
\end{eqnarray}
In the quantum mechanics approach, the transition from $
\bar{D}^*D/\bar{D}D^*$ to $J/\psi \omega(\rho)$ occurs in  very
short-range region, because the heavy (anti) quarks have to be
reclustered. The transition amplitude is expected to be proportional
to the molecular wave function at the origin and  reads,
\begin{eqnarray}
    {\cal A}^{I=a}  [X\to J/\psi V]&\sim&\sum_{i=1,2}c_{ai}\varphi_{i}(0) \nonumber\\
    &\sim&c_{a1}{G}_{1}{g}_{1}+c_{a2}{G}_{2}{g}_{2},
\end{eqnarray}
where $\varphi_i(0)$ are the wave functions at the origin  in
Eq.~\eqref{eq:wv_orgin}. With the above two approaches, we get the
same results.

Substituting the specific Clebsch-Gordan coefficients, the ratio
$R_1$ can be calculated as
\begin{eqnarray}
    R_{1}&=&\left(\frac{{g}_{1}{G}_{1}-{g}_{2}{G}_{2}}{{g}_{1}{G}_{1}+{g}_{2}{G}_{2}}\right)^{2}\approx\left(\frac{{g}_{1}-{g}_{2}}{{g}_{1}+{g}_{2}}\right)^{2} \nonumber\\
    &=&\left(\frac{{c}_{1}{\kappa}_{1}^{1/2}-{c}_{2}{\kappa}_{2}^{1/2}}{{c}_{1}{\kappa}^{1/2}+{c}_{2}{\kappa}^{1/2}}\right)^{2}. \label{eq:R1}
\end{eqnarray}
Since our framework is independent on the cutoff parameter, one can
take the $\Lambda \gg {\kappa}_i$ and then can eliminate the ${G}_i$
according to Eq.~\eqref{eq:Gi}.

In Ref.~\cite{Gamermann:2009fv}, the authors presumed ${g}_1={g}_2$.
They had to keep the difference of ${G}_1$ and ${G}_2$ to make $R_1$
non-vanishing, which is a cutoff-dependent result. In our approach,
with a general interaction~\eqref{eq:vij}, the cutoff dependence can
be canceled out when $\Lambda \gg \kappa_i$.  Without constraint of
$v_{11}=v_{22}=v_{12}$ in Ref.~\cite{Gamermann:2009fv}, the $g_i$ in
Eq.~\eqref{eq:g1g2} is proportional to the $c_i$ and $\kappa_i$. The
coupling constants are determined by the mixing angle of two
channels, binding energy and mass differences of two thresholds,
which are all physical observables. Therefore, the results in
Ref.~\cite{Gamermann:2009fv} kept the cutoff-dependent effect from
${G}_1$ and ${G}_2$ and neglected the more physical effect from
$g_1$ and $g_2$. In fact, the coupling constants embed important
information about the structure of $X(3872)$. In our approach, the
drawback of Ref.~\cite{Gamermann:2009fv} is cured by introducing
more general interaction in Eq.~\eqref{eq:vij} in a framework
satisfying the renormalization group invariance.

\begin{figure}[!htp]
    \includegraphics[width=0.4\textwidth]{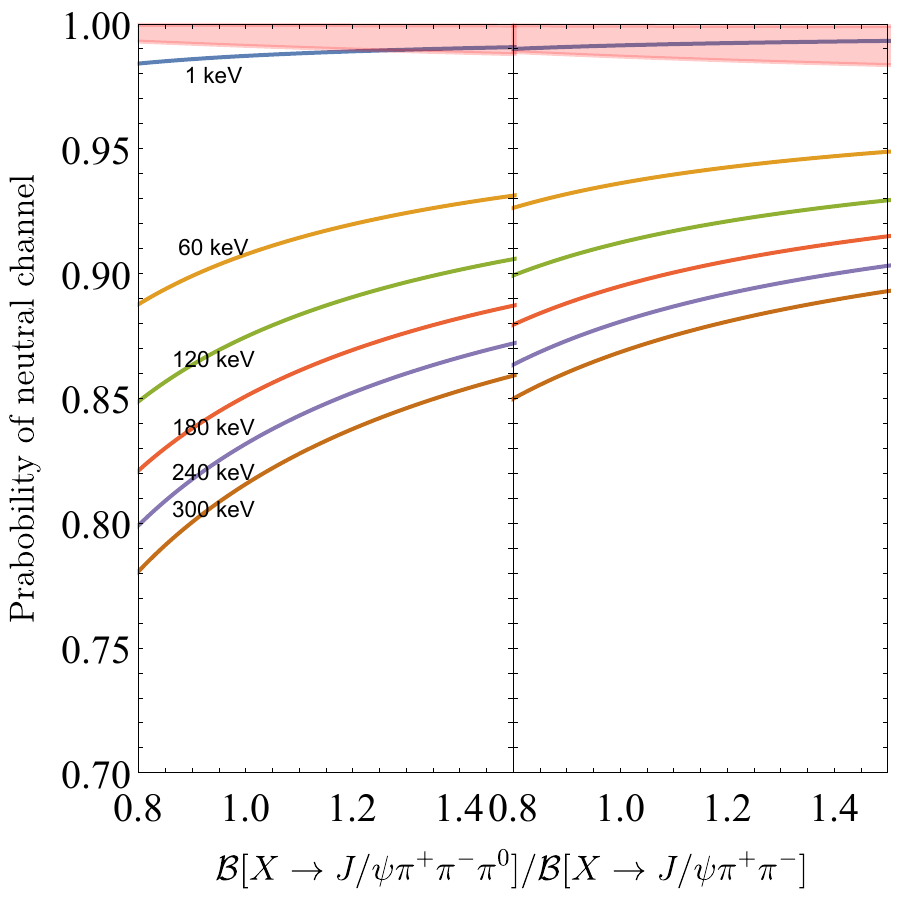}
    \caption{The probability of the neutral channel in the $X(3872)$ wave function. In the left and right subfigures, we adopt different $R_1$ from our calculation and Ref.~\cite{Braaten:2005ai}, respectively. We vary the binding energy of $X(3872)$ from $300$ to 1 keV. There are two solutions in each set of input. The first solutions are very close to 1, which are represented by the red shadow. The second solutions are represented by solid lines with the corresponding binding energy near it. }\label{fig:xmixing}
\end{figure}

With the $R_2$ and $R_1$ in Eqs.~\eqref{eq:R2} and~\eqref{eq:R1}, we
can extract the probabilities (${c}_1^2$ and ${c}_2^2$) of the
neutral and the charged channels from the experimental $R$. Since
the binding energy is unknown, we vary it from 300 keV to 1 keV. In
addition to our value of  $R_2=0.147$, we also use the value
$R_2=0.087$ in Ref.~\cite{Braaten:2005ai}. By varying the
experimental $R$ from 0.8 to
1.5~\cite{Belle:2005lfc,BaBar:2010wfc,BESIII:2013fnz}, we obtain two
solutions of $\tan{\theta}= {c}_2/{c}_1$ for every parameter set.
Both solutions are in the range of (0,1/2). Correspondingly, the
mixing angle is in the range of $(0,\pi/4)$. The ${c}_1^2$ that
extracted from the two solutions are presented in
Fig.~\ref{fig:xmixing}. The ${c}_1^2$ for the first solution is very
close to 1. The second solution of ${c}_1^2$ is also larger than
80\%, which increases with  decreasing the binding energy. Thus,
$X(3872)$ is a bound state dominated by the neutral channel, which
has a proportion larger than 80\%. In Ref.~\cite{Wu:2021udi}, the calculations in a cutoff-dependent scheme indicated that in the $X(3872)$, the weight of the neutral component is $(83-88)\%$, which agrees with our results.

In Ref.~\cite{Li:2012cs}, the authors obtained a result equivalent
to $R_1=({c}_1-{c}_2)^2/({c}_1+{c}_2)^2$ in our notation.  The
amplitude $X(3872)\rightarrow J/\psi V$ is assumed to be
proportional to the coefficients of the related components rather
than the wave functions at the origin. The proportions of the
neutral channel in Ref.~\cite{Li:2012cs} is close to our
calculations. However, transition from a $\bar{D}^*D/\bar{D}D^*$
bound state to the $J/\psi V$ channel is a short-range process.
Taking the ratio of components of wave functions may be still a
good approximation for $R_1$. But using the wave functions at the
origin is more reasonable.

In our previous work~\cite{Meng:2020cbk}, we considered the $D_s \bar{D}^*_s / D^*_s\bar{D}_s $ systems by relating their interaction to the $D \bar{D}^* / D^*\bar{D} $ one. But we neglected the coupled-channel effect between them for the large mass splitting $\delta_{s}=D_s^{(*)}-D^{(*)}\approx 100$ MeV. $X(3872)$ is in the proximity of the $D \bar{D}^* / D^*\bar{D} $  about several MeVs, which are much smaller than the mass splitting $\delta_s$. Thus, it is reasonable to neglect the $D_s \bar{D}^*_s / D^*_s\bar{D}_s $ channel. In Ref.~\cite{Gamermann:2009fv}, the authors took the $D_s \bar{D}^*_s / D^*_s\bar{D}_s $ into consideration in the local hidden gauge approach. The coupling constants were obtained from the residues of the $T$-matrix, which implied a sizable fraction of $D_s \bar{D}^*_s / D^*_s\bar{D}_s $ component. However, one should notice that the conclusion in Ref.~\cite{Gamermann:2009fv} is model-dependent. The di-meson interaction derived from the local hidden gauge approach is very strong, which makes the mass splitting $\delta_{s}$ play a minor role. Thus, the SU(3) flavor breaking effect is not significant.  Meanwhile, the loop diagrams are regularized in dimensional regularization scheme with a varying subtraction coefficient. The solution of $X(3872)$ can be reproduced by tuning the subtraction coefficient. One can reproduce the pole of $X(3872)$ without the $D_s \bar{D}^*_s / D^*_s\bar{D}_s $ channel by choosing a different subtraction coefficient. The coefficient is equivalent to the cutoff parameter in the cut-off regularization scheme. Thus, the conclusion in Ref.~\cite{Gamermann:2009fv} is cutoff-dependent. 

In Ref.~\cite{Aceti:2012cb}, the authors specified processes $\bar{D}^*D/\bar{D}D^*\to J/\psi \rho(\omega)$ [``black box'' of Fig.~\ref{fig:feynjpsipipi}] by exchanging explicit particles. In principle, the isospin symmetry taken in this work would be a good approximation  in the specific mechanism of Ref.~\cite{Aceti:2012cb} as well. One can expect qualitatively consistent conclusions with this work if the similar coupling constants were taken.  However, the authors took the the coupling constants in Ref.~\cite{Gamermann:2009fv}, which included sizable $D_s \bar{D}^*_s / D^*_s\bar{D}_s $ components.  The comparable amplitudes of $D_s \bar{D}^*_s / D^*_s\bar{D}_s \to J/\psi \rho$ and  $D_s \bar{D}^*_s / D^*_s\bar{D}_s \to J/\psi \omega$ will give the considerable isospin violation effect. We should stress that the large proportion of $D_s \bar{D}^*_s / D^*_s\bar{D}_s$ in $X(3872)$ is just a model-dependent result.

\begin{figure*}[!htp]
    \centering  \includegraphics[width=0.8\textwidth]{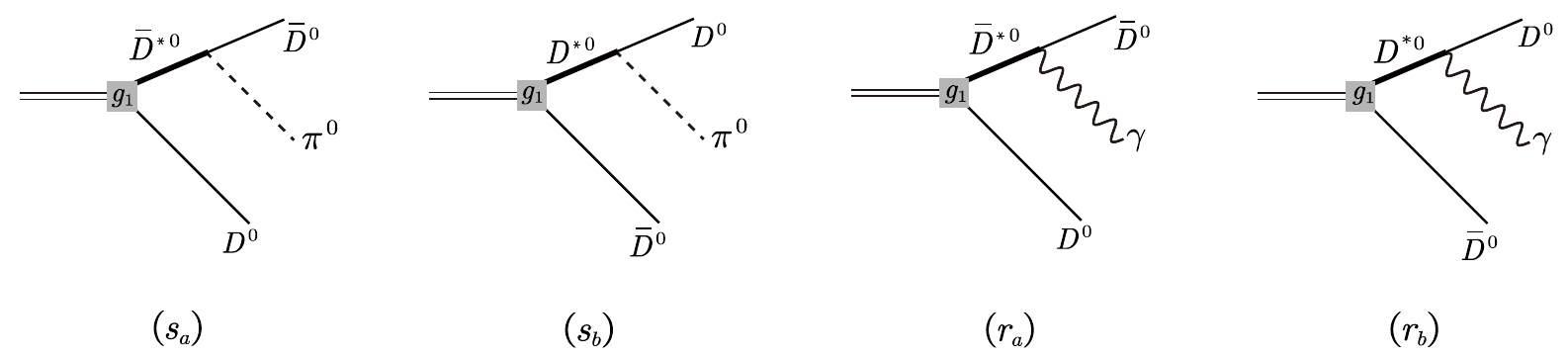}
    \caption{The Feynman diagrams for strong and radiative decays of the $X(3872)$ state, where only the neutral channel contributions are illustrated.}\label{fig:feynx}
\end{figure*}

\section{Strong and radiative decays of $X(3872)$}~\label{sec:decay}

With the mixing angle $\theta$, we can calculate the strong decay
$X(3872)\to \bar{D}^0 D^0\pi^0$ and radiative decay $X(3872)\to
\bar{D}^0 D^0\gamma$. In Ref.~\cite{Guo:2014hqa}, the authors
calculated the two decay widths with the similar interactions in
Eq.~\eqref{eq:vij}, where the low energy  constants are determined
by the invariant mass distributions of the $J/\psi \rho (\omega)$
final states~\cite{Hanhart:2011tn} and mass of
$Z_b(10610)$~\cite{Cleven:2011gp}. In our calculation, we only
consider the contribution from the tree diagrams as shown in
Fig.~\ref{fig:feynx}. The coupling constants  between $X(3872)$ and
$\bar{D}^*D/\bar{D}D^*$ are related to the binding energy and the
mixing angle. In this work, we vary the probability of the neutral
component of $X(3872)$ from 80\% to 100\% and  the binding energy
from 300 keV to 0 keV, respectively.  The vertex $D^*\to D\pi$ and
$D^*\to D\gamma$ are extracted from the experimental widths of
$D^*$.

The strong decay $X(3872)\to \bar{D}^0 D^0\pi^0$ is induced by the
intermediate neutral ${\bar D^{*0}}D/D^{*0}\bar D$ channel as shown
in Fig.~\ref{fig:feynx}. The decay vertex $D^{\ast0}\to D^0\pi^0$ is
determined by  $D^{\ast+}\to D^0\pi^+$ using the isospin symmetry
due to the lack of experimental  $D^{\ast0}$ width. The amplitude of
the decay $D^{\ast+}\to D^0\pi^+$  is $\mathcal{A}=g_\pi q_\pi \cdot
\epsilon_{D^*}$, where $\epsilon_{D^*}$ and $q_\pi$ are the
polarization vector of $D^*$ meson and momentum of pion,
respectively. We extract the coupling constant $g_\pi=11.25$ as our
input.

For the radiative decay, we only list the decay mode $X(3872)\to
\bar{D}^0D^0\gamma$  in Fig.~\ref{fig:feynx}, which is driven by the
neutral channel. Another radiative decay mode $X(3872)\to
D^+D^-\gamma$  should be largely suppressed. On the one hand, it is
driven by the charged channel, which only occupies very minor
proportion, less than 20\%. On the other hand, the leading
amplitudes for M1 radiative transitions $D^{*0,+}\to D^{0,+}\gamma$
are roughly proportional to the electric charges of the light quarks
in the heavy quark limit~\cite{Wang:2019mhm}. The $D^{*+}\to
D^+\gamma $ is suppressed compared with $D^{*0}\to D^0\gamma$. Thus,
the partial decay width of the $X(3872)\to D^+D^-\gamma$ is very
tiny. The radiative decay vertex of $D^{*}\to D\gamma$ can be
parameterized as follows,
\begin{eqnarray}
    {\cal A}[D^{*}\to D\gamma]=g_{\gamma}\varepsilon_{\mu\nu\alpha\beta}\epsilon_{\gamma}^{\mu}p_{D^{*}}^{\nu}p_{\gamma}^{\alpha}\epsilon_{D^{*}}^{\beta},
\end{eqnarray}
where $g_\gamma$ denotes the effective coupling constant. Its value
is extracted from the partial decay width of $D^{\ast 0}\to
D^{0}\gamma$. For the $D^{\ast0}$ meson, we take its total width as
$60$ keV, which approaches to most of the theoretical results,
e.g.~\cite{Ebert:2002xz,Choi:2007se,Becirevic:2009xp,Wang:2019mhm}.

\begin{figure}[!htp]
    \includegraphics[width=0.4\textwidth]{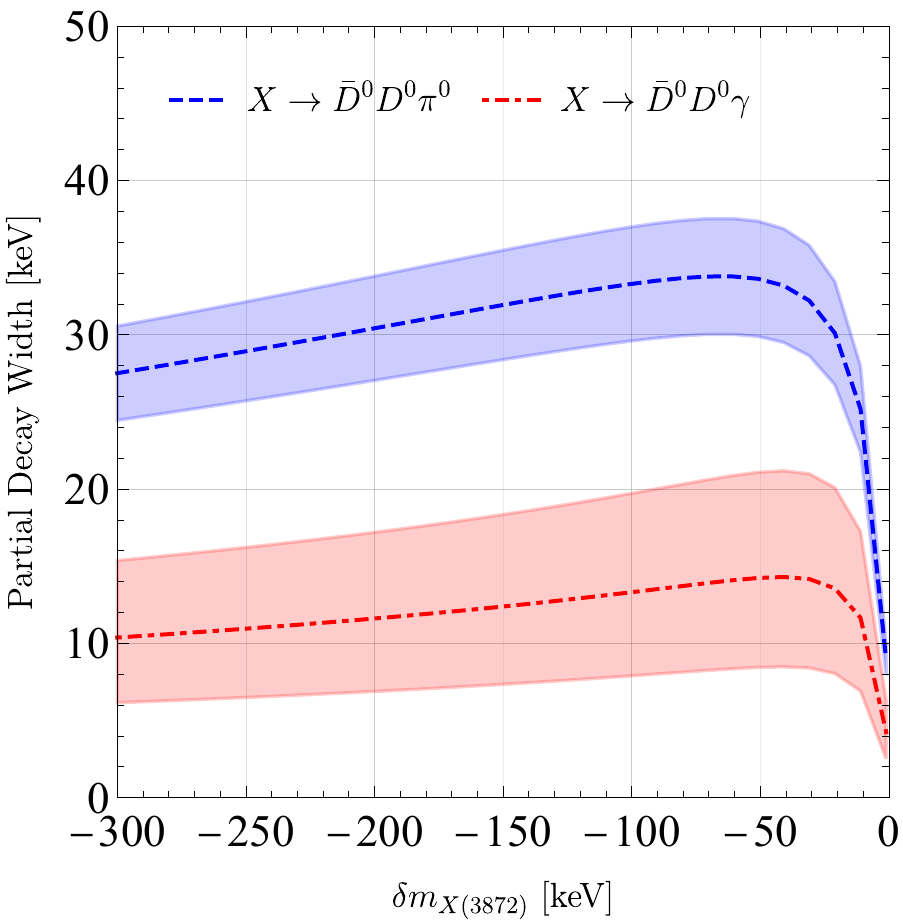}
    \caption{The partial decay widths of $X(3872)$ with its binding energy. The shadows represent the probability of the neutral channel in the range (80\%, 100\%). The central values corresponding to results with 90\% neutral component.  }\label{fig:xdecay}
\end{figure}

In Fig.~\ref{fig:xdecay}, we present the numerical results of the
strong decay $X(3872)\to \bar{D}^0 D^0\pi^0$ and radiative one
$X(3872)\to \bar{D}^0 D^0\gamma$. The results show that the strong
and radiative decay widths are about 30 keV and 10 keV,
respectively, when the binding energy is in the range  of (-300,
-50) keV. Both decay widths keep increasing with the larger
$X(3872)$ mass (larger phase space) until the binding energy is
about $-50$ keV. When the $X(3872)$ approaches the neutral
threshold, the coupling constant $g_1$ will decrease significantly
since it is proportional to the binding momentum. The strong and
radiative decay widths then tend to vanish because of small $g_1$
even with larger phase space.

\section{Summary}~\label{sec:sum}
In this work, we revisit the isospin violating decays of $X(3872)$
with a coupled-channel effective field theory. In the molecular
scheme, the $X(3872)$ is interpreted as the bound state of
$\bar{D}^{*0}D^0/\bar{D}^0D^{*0}$ and $D^{*-}D^+/D^-D^{*+}$
channels. In a cutoff-independent formalism, we relate the coupling
constants of  $X(3872)$ to  two channels of its wave function. The
isospin violating decays of $X(3872)$ are obtained by two equivalent
approaches, which ameliorate the approaches for this issue in
literature. In the quantum field theory approach, the isospin
violating decays arise from the coupling constants of $X(3872)$ to
two di-meson channels. We show that the driving source of the
isospin violating proposed in Ref.~\cite{Gamermann:2009fv} is
cutoff-dependent. However, the factors (coupling constants) related
to the physical observables were neglected. In the quantum mechanics
approach, the isospin violating is attributed to wave functions at
the origin, which is also different from that in
Ref.~\cite{Li:2012cs} focusing on the coefficients of different
channels. The transition from a $\bar{D}^*D/\bar{D}D^*$ bound state
to the $J/\psi V$ is a short-range process. Therefore, using the
wave function at the origin is more reasonable.

Within a comprehensive analysis, we bridge the isospin violating
decays of $X(3872)$ to its structure. Our results show that the
proportion of the neutral channel in $X(3872)$ is over $80\%$. As a
by-product, we calculate the strong decay width of $X(3872)\to
\bar{D}^0 D^0\pi^0$ and radiative decay width $X(3872)\to \bar{D}^0
D^0\gamma$. The strong decay width and radiative decay width are
about 30 keV and 10 keV, respectively, for the binding energy from
$-300$ keV to $-50$ keV.

The isospin violating decays of $X(3872)$ were investigated in
amounts of literature. In this work, we do not aim to provide very
refined calculations but try to improve some conceptional
understandings. $X(3872)$ is a benchmark of the exotic hadrons. We
hope our work could shed some light on its structure. The approach
in Ref.~\cite{Gamermann:2009fv} was also adopted to detect the
molecular structure of $P_c(4457)$~\cite{Guo:2019fdo}. The authors
neglected the isospin violating effect in coupling constants but
drive the isospin violation by the mass splitting in propagators.
The cutoff-dependence in the final results in
Ref.~\cite{Guo:2019fdo} can also be improved by the approach in this
work.

\begin{appendix}

\section{Sequential decay}~\label{app:sq_decay}
We use the  decay process  $A\to ab\to a(123)$  where $A$ decays
into $a$ and $b$, and then  $b$ sequentially decays into three
particles to  illustrate the calculation of sequential decay.  For
simplification, we  presume $b$ as a (pseudo)scalar particle and
the decay amplitude reads,
\begin{equation}
    {\cal T}={\cal A}_{A}[A\to ab]\frac{i}{p_{b}^{2}-m_{b}^{2}+im_{b}\Gamma_{b}}{\cal A}_{b}[b\to123],\label{eq:t_14}
\end{equation}
where $\mathcal{A}_A$ and $\mathcal{A}_b$ are the decay amplitudes
in sequence. The $p_b$, $m_b$ and $\Gamma_b$ are the momentum, mass
and width of particle $b$, respectively.  The differential decay
width reads,
\begin{widetext}
\begin{eqnarray}
    d\Gamma&=&\frac{(2\pi)^{4}}{2M_{A}}|{\cal T}|^{2}d\Phi_{4} \nonumber\\
    &=&\left[\frac{(2\pi)^{4}}{2M_{A}}|{\cal A}_{A}|^{2}d\Phi_{a}\right]\times\left[\frac{m_{b}}{(p_{b}^{2}-m_{b}^{2})^{2}+m_{b}^{2}\Gamma_{b}^{2}}\frac{1}{\pi}dp_{b}^{2}\right]\times\left[\frac{(2\pi)^{4}}{2m_{b}}|{\cal A}_{b}|^{2}d\Phi_{123}\right] \nonumber\\
    &\approx&\left[\frac{(2\pi)^{4}}{2M_{A}}|{\cal A}_{A}|^{2}d\Phi_{a}\right]\times\left[\frac{m_{b}}{(p_{b}^{2}-m_{b}^{2})^{2}+m_{b}^{2}\Gamma_{b}^{2}}\frac{1}{\pi}dp_{b}^{2}\right]\times\Gamma_{b\to123}\nonumber\\
    &=&\left[\frac{(2\pi)^{4}}{2M_{A}}|{\cal A}_{A}|^{2}d\Phi_{a}\right]\times\left[\frac{m_{b}\Gamma_{b}}{(p_{b}^{2}-m_{b}^{2})^{2}+m_{b}^{2}\Gamma_{b}^{2}}\frac{1}{\pi}dp_{b}^{2}\right]\times{\cal B}_{b\to123},~\label{eq:diff_wd}
\end{eqnarray}
\end{widetext}
with the general $n$-body phase space as
\begin{equation}
    d\Phi_{n}(P;p_{i},...p_{n})=\delta^{4}(P-\sum_{i}^{n}p_{i})\prod_{i}^{n}\frac{d^{3}p_{i}}{(2\pi)^{3}2E_{i}}.
\end{equation}
$P$ and $p_i$ are the momenta of the inital and final particles,
respectively. The $4$-body phase space in Eq.~\eqref{eq:diff_wd} is
decomposed as follows,
\begin{equation}
    d\Phi_{4}=d\Phi_{a}(P;p_{b},p_{a})d\Phi_{123}(p_{b};p_{1,}p_{2},p_{3})(2\pi)^{3}dp_{b}^{2}.
\end{equation}
In Eq.~\eqref{eq:diff_wd}, the $\mathcal{A}_b$ and $\Phi_{123}$ both
depend on $p_b^2$. However, for the almost on-shell $b$, it will be
a good approximation to replace the $p_b^2$ with $m_b^2$. Therefore,
one can factor out the $\Gamma_{b\to 123}$ in
Eq.~\eqref{eq:diff_wd}. The above derivation can be easily extended
to the sequential decay $A\to ab\to a(12)$ with $b$ being a vector
particle. As shown in Eq.~\eqref{eq:r2}, the $\omega \to
\pi^+\pi^-\pi^0 (\rho^0\to \pi^+\pi^-)$ decay widths are factored
out and the polarizations of the particles are neglected.

\end{appendix}

%\vspace{0.5cm}

\begin{acknowledgements}
We are grateful to the helpful discussions with Prof.~Eulogio Oset.
L. M. is grateful to the helpful communications with Prof.~Dian-Yong
Chen. This project was supported by the National Natural Science
Foundation of China (11975033 and 12070131001). This project was
also funded by the Deutsche Forschungsgemeinschaft (DFG, German
Research Foundation, Project ID 196253076-TRR 110). G.J. Wang was
supported by JSPS KAKENHI (No.20F20026). B. Wang is supported by the Youth Funds of Hebei Province (No. 042000521062) and the Start-up Funds for Young Talents of Hebei University (No. 521100221021).
\end{acknowledgements}

\bibliography{x3872}

\end{document}